\documentstyle[emulateapj,pstricks,psfig]{article}

\def\\{\hfill\break}

\def\etal{{et al.\ }}

\def\be{\begin{equation}}
\def\ee{\end{equation}}

\def\ifm#1{\relax\ifmmode#1\else$\mathsurround=0pt #1$\fi}
\def\kms{\ifmmode\,{\rm km}\,{\rm s}^{-1}\else km$\,$s$^{-1}$\fi}
\def\hmpc{\,\ifm{h^{-1}}{\rm Mpc}}
\def\hkpc{\,{\rm h^{-1}kpc}}

\def\dd{{\rm d}}

\def\msolar{M_{\odot}}
\def\hmsun{h^{-1}\msolar}
\def\hMsun{\hmsun}
\def\ltsima{$\; \buildrel < \over \sim \;$}
\def\lsim{\lower.5ex\hbox{\ltsima}}
\def\gtsima{$\; \buildrel > \over \sim \;$}
\def\gsim{\lower.5ex\hbox{\gtsima}}

\def\pmb#1{\setbox0=\hbox{#1}%
 \kern-.025em\copy0\kern-\wd0
 \kern.05em\copy0\kern-\wd0
 \kern-.025em\raise.0433em\box0}

\def\v0{\pmb{$0$}}

\def\rs{r_{\rm s}}
\def\rhos{\rho_{\rm s}}
\def\rvir{r_{\rm v}}
\def\mvir{m_{\rm v}}

\def\vvir{v_{\rm v}}
\def\vmax{v_{\rm m}}
\def\vm{v_{\rm m}}
\def\rmax{r_{\rm m}} 

\def\tbe{\tilde\beta}
\def\tz{\tilde z}

\def\lcdm{$\Lambda$CDM}
\def\omm{\Omega_{\rm m}}
\def\oml{\Omega_{\Lambda}}

\begin{document}
\slugcomment{{\em Astrophysical Journal Letters, submitted}}

\lefthead{EVOLUTION IN THE LUMINOSITY-VELOCITY RELATION}
\righthead{BULLOCK, DEKEL, PRIMACK, \& SOMERVILLE}
\title{Strong Evolution in the Luminosity-Velocity Relation at $z \ga 1$?}
\author{James S. Bullock \altaffilmark{1}}
\affil {Department of Astronomy, Ohio State University,
Columbus, OH 43210}

\author {Avishai Dekel \altaffilmark{2}}
\affil {Racah Institute of Physics, The  Hebrew University,
Jerusalem 91904, Israel}

\author {Joel R. Primack \altaffilmark{3}}
\affil {Physics Department, University of California, Santa
Cruz, CA 95064}

\author {Rachel S. Somerville \altaffilmark{4}}
\affil {Institute of Astronomy, Madingley Rd., Cambridge CB3 OHA, UK}

\altaffiltext{1}{james@astronomy.ohio-state.edu}
\altaffiltext{2}{dekel@astro.huji.ac.il}
\altaffiltext{3}{joel@ucolick.org}
\altaffiltext{4}{rachel@ast.cam.ac.uk}


\begin{abstract}


We present a  method  for  constraining  the evolution  of the  galaxy
luminosity-velocity  (LV)   relation  in  hierarchical   scenarios  of
structure formation. The comoving  number density of dark-matter halos
with circular   velocity  of $200\kms$  is  predicted  in  favored CDM
cosmologies   to be nearly constant   over  the redshift range $0\lsim
z\lsim  5$.  Any observed evolution in  the density of bright galaxies
implies in turn a corresponding   evolution in the  LV relation.    We
consider several possible forms of evolution for the zero-point of the
LV relation and predict  the corresponding evolution in galaxy  number
density.  The  Hubble Deep Field suggests   a  large deficit of bright
($M_V<-19$) galaxies at $1.4\lsim z\lsim  2$.  If taken at face value,
this implies a  dimming of the LV zero-point  by roughly 2 magnitudes.
Deep, wide-field, near-IR selected surveys   will provide more  secure
measurements to compare with our predictions.

\end{abstract}

\subjectheadings{cosmology: theory --- dark matter --- galaxies:
formation --- large-scale structure of universe}

\section{Introduction}
\label{sec:intro}

Within the  framework  of  hierarchical structure  formation, galaxies
form within virialized dark-matter (DM) halos. The halo properties and
their time  evolution can be  predicted  robustly within  a given Cold
Dark  Matter  (CDM)   model  using  N-body  simulations    or analytic
approximations.   A fundamental property of  a  DM halo is its maximum
circular velocity,  which is likely  to  be related to  the observable
internal velocity of the luminous galaxy that resides in the halo.

In the local universe, it is well known that galaxies obey fundamental
scaling relations, including the relation between total luminosity and
internal velocity  (commonly  known  as Tully-Fisher or  Faber-Jackson
relations), which we will focus  on here.  This LV relationship surely
reflects some fundamental  aspect  of galaxy formation,   although its
origin is poorly understood.  An observational determination of the
\emph{redshift evolution} of  this relationship would  provide crucial
clues as  to its physical  origin  and  would constitute an  important
constraint on models  of     galaxy formation.  However,      a direct
observational determination of  the internal velocities of galaxies at
high  redshift is  extremely  difficult.   Rotation  curves have  been
obtained for a small number of galaxies up  to redshift $z\sim1$ (Vogt
et al. 1996, 1997, 2000).  These results  suggest that there is little
evolution  in the Tully-Fisher relation for  objects  similar to local
large spirals out  to $z\sim1$ (but see Mallen-Ornelas  et al.  1999).
At higher   redshifts  ($z\ga 1$),   no   direct observations  are yet
available.

While  galaxies  of different  morphological types   obey different LV
relations, one  can   define  a   mean LV   relation for   any   given
morphological  mix.   This can  be used   to  make a direct connection
between the distribution  functions of   galaxies  in terms  of  their
\emph{internal velocity} and
\emph{luminosity}. 
Clearly, if  the evolution  of the  velocity function  is predicted by
theory, and  the evolution  of  the luminosity function  is determined
observationally,  this   connection can provide   a  constraint on the
evolution of the LV relation.

Determining the  number density  of galaxies as  a function   of their
rest-frame luminosity  is  considerably  easier  than  measuring their
rotation velocities directly at  high redshift. Observed  luminosities
can be k-corrected  to  the rest visual   frame up  to $z\sim2$  using
multi-band photometry  extending  into the  near-IR. It has  been very
difficult to obtain spectroscopic redshifts  in the `desert' regime of
$1.4 \lsim z  \lsim  2$, but  photometric  redshifts offer a  means of
filling   in this gap.   These    advances, combined with  theoretical
progress in predicting the number density and structural properties of
dark matter halos, suggest the approach  for constraining the redshift
evolution of the LV relation that we exploit here.

The basis of our analysis is the evolution of the velocity function of
DM halos as determined via a high-resolution N-body simulation of the
\lcdm  cosmology (Sigad et al.  2000, hereafter S00).  We extend these
results to  other cosmologies using   analytic models.  We  consider a
wide  range of  possibilities for the   redshift evolution  of  the LV
relation,  including some motivated  by  the results of  semi-analytic
models (cf. Somerville \&  Primack 1999).  We then present predictions
of the corresponding redshift evolution of the comoving number density
of bright  galaxies, which may be  compared directly with observations
to provide quantitative constraints on how the  LV relation evolves in
time.

\section{Velocity Function}
\label{sec:vf_lcdm}

We have calculated the  velocity function of  DM halos using an N-body
simulation  of   the currently    popular    flat \lcdm\  model   with
$\Omega_{\rm m}  =0.3$,  $\Omega_{\rm \Lambda} =0.7$,  Hubble constant
$h=0.7$,  and $\sigma_{8} =  1.0$.  The simulation (Klypin \etal 1999)
used a high-resolution adaptive refinement tree code (Kravtsov, Klypin
\& Khokhlov 1997) to follow the evolution of structure in a periodic
box of comoving side 60$\hmpc$, with a force resolution of $\sim
2\hkpc$ in the densest regions.

We   identify     bound, virialized     halos    using  a    spherical
overdensity-based  algorithm and  the  standard definition  of  virial
density from  the spherical top-hat  collapse model (cf. Eke, Cole, \&
Frenk 1996;  Bryan  \& Norman 1999). Details   of the halo  finder are
described in Bullock \etal (2000;  hereafter B00). Although the virial
velocity  $\vvir  \equiv [G   \mvir/\rvir]^{1/2}$ (where  $\mvir$  and
$\rvir$ ar the virial mass and radius)  is often used to approximate
galactic circular velocities, a more relevant quantity observationally
is the maximum  rotation velocity of   the halo $\vmax$, which can  be
related to $\vvir$ via the halo  density profile. Over relevant radii,
halo density profiles are well approximated by the NFW functional form
with two free parameters (Navarro, Frenk \& White 1996): $\rho(r)=
\rhos (r/\rs)^{-1} (1+r/\rs)^{-2}$, where $\rs$ is the radius at which
the   log slope of  the profile  is $-2$.   The  halo maximum circular
velocity  occurs  at $2.15\rs$  and  is  related  to  $\vvir$ via $\vm
=\sqrt{c/f(c)}\vvir$, where   $f(c)=4.62    [\ln (1+c) -c(1+c)^{-1}]$.
Here,  $c$  is the  ``concentration''  parameter $c=\rvir  /\rs$.  The
value of $c$ is closely related to the halo mass, and for a typical DM
halo surrounding a bright galaxy ($\mvir \sim 10^{12}
\hMsun$) at $z=0$, our simulations yield $c \sim 13$, implying $\vm =
1.3 \vvir$.   The concentration at a  fixed mass evolves  rapidly with
redshift, $c  \propto  (1+z)^{-1}$   (B00), implying  a  corresponding
evolution in $\vm(\vvir)$.  The halo catalogs described in B00 provide
$\vm$ and  $\vvir$  for  each halo.  A   detailed  study of  the  halo
velocity function derived from these catalogs is presented by S00.

{\pspicture(0,0)(10,10.)
\rput[tl]{0}(-.2,10.){\epsfxsize=9.cm
\epsffile{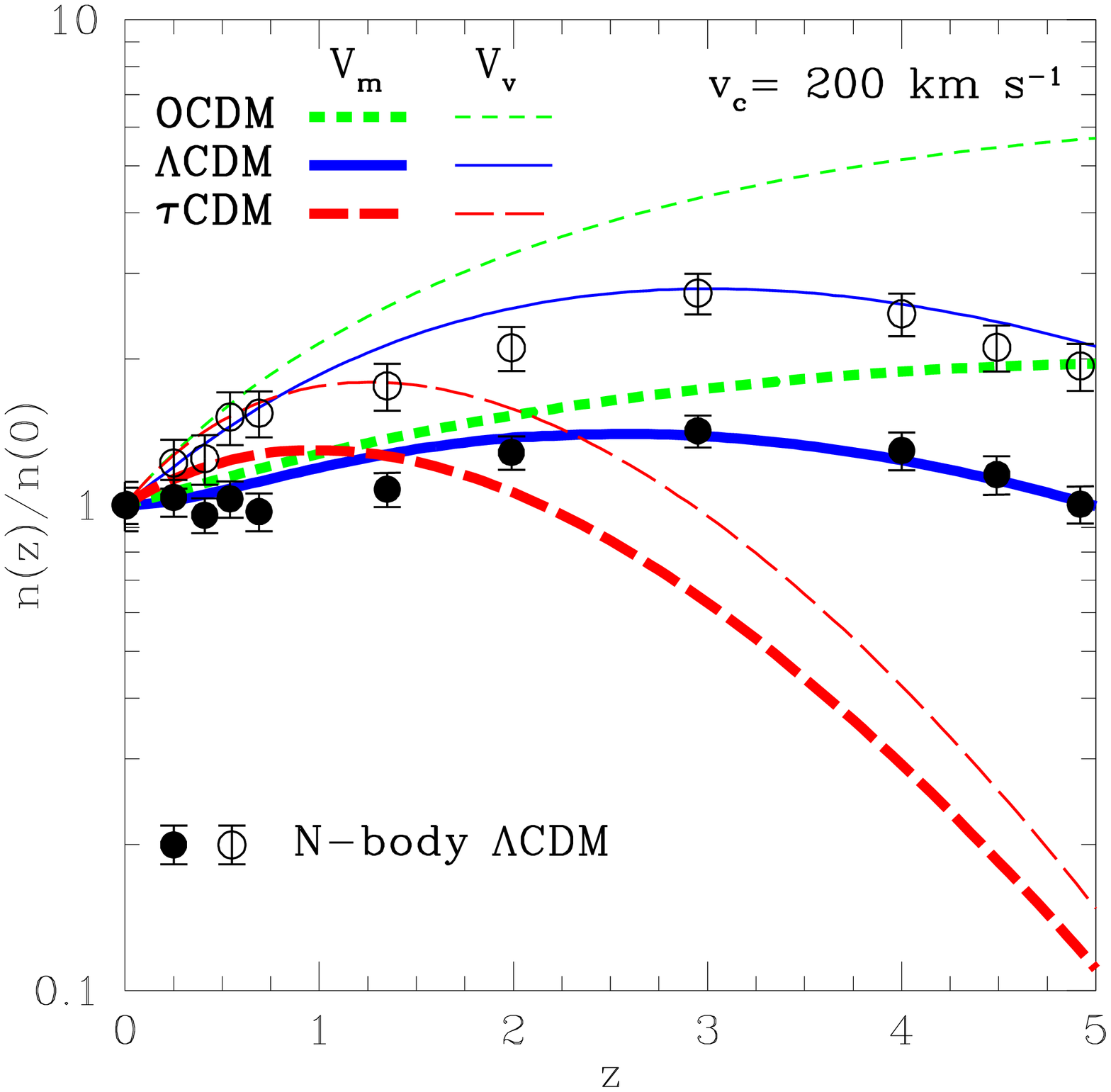}}
\rput[tl]{0}(0,1.){
\begin{minipage}{8.5cm}
  \small\parindent=4.5mm   {\sc  Fig.}~1.---  Evolution   of  relative
  comoving number density for fixed $\vm =200  \kms$ (bold curves) and
  $\vvir = 200 \kms$ halos in three cosmologies.
\end{minipage}
}
\endpspicture}

Figure 1 shows  the  evolution with  redshift of  the comoving  number
density of simulated halos of  fixed velocity, either $\vm=200\kms$ or
$\vvir=200\kms$.  The evolution of halos of fixed $\vm$ is weaker than
for halos of fixed $\vvir$ due to the strong evolution of $c$.  A fit
\be
\dd  n= \psi(\vm)\,\dd (\log \vm) =\psi_* \vm^{\tbe} \,\dd (\log \vm) \ \ ,
\label{eq:vfit_log}
\ee
for the whole halo $\vm$ distribution is provided by 
S00, who find that the comoving density at $\vm=300\kms$ is constant as a
function of $z$ ($\pm 0.1$ in the log) at least out to $z=5$, while
$\tbe \approx -2.9 -2.0\tz + 1.4\tz^2 \pm 0.1$,  where $\tz\equiv\log(1+z)$.

Figure 1 also shows predictions based on the analytic approximation of
Sheth \&   Tormen  (1999, hereafter ST)  ---    a modification  of the
Press-Schechter approximation (1974) which improves the agreement with
simulations.  We  converted the ST mass  to $\vvir$  and then to $\vm$
using the  analytic model of  B00 for  $c(\mvir,z)$.  We see  that for
\lcdm,  the analytic  model  provides  a  good  approximation  to  the
simulated evolution  both  for fixed  $\vm$  and fixed  $\vvir$.  This
allows us to extend our predictions using ST to two other cosmological
models:  $\tau$CDM, a flat model   with $\omm=1$, $h=0.5$ and a  power
spectrum of shape parameter $\Gamma=0.21$,  and OCDM, an open geometry
with $\omm=0.3$, $\oml=0$, and $h=0.7$.

In the case  of  \lcdm, we note  an  increase of  $\sim 30\%$   in the
density for constant $\vm$ between $z=0$ and  $z\sim 3$, and a gradual
decline back to today's density by $z\sim  5$. The density of halos of
a constant $\vvir$ grows by a factor $\sim 2.6$ by $z\sim 3$, and then
declines at higher $z$.   Note that a halo with  a fixed $\vvir$ has a
decreasing mass and virial radius at increasing redshift.  Compared to
\lcdm, the OCDM model predicts a slightly larger increase in the density of
$200\kms$ halos at all redshifts.  In the case  of $\tau$CDM, the peak
density is obtained at $z\sim  1$, followed by  a sharper decline, but
note that even  in  this case the  comoving density  at $z\sim   2$ is
comparable to the density at $z=0$.

\section{LV relations and luminosity functions}
\label{sec:lfun}

Armed with this knowledge  about  the evolution  of the  halo velocity
function, we can  now predict the evolution  of the number  density of
galaxies brighter than a  fixed absolute luminosity  for a given model
of LV evolution.

A  possible hitch is  that the halo  maximum  velocity $\vm$ typically
occurs at a radius $\rmax$ that lies outside the luminous parts of the
galaxy for which  the  LV quantities are    measured. For example,   a
$\vm=200 \kms$ halo at $z=0$ has $\rmax \sim 45$kpc.  The velocity due
to the   halo alone obviously   declines at  smaller radii,   but  the
condensed baryons  contribute significantly   to  the total   circular
velocity there.   We therefore make  the  assumption that the rotation
curve  is roughly flat between where  it is measured and $\rmax$, i.e.
that the halo $\vm$ is the same  as the velocity  used in the observed
LV  relations.  Gonzalez \etal    (1999; G99)\footnote{ The  published
version of  this manuscript contains an  error in the Equation 12: the
quantity in the square root should be inverted.}  used this assumption
at  $z=0$  to  compare the velocity  function  of  DM halos  with that
obtained from observed luminosity functions and LV relations. The good
agreement that  they  obtained is  evidence that this  is a reasonable
approximation at low redshift. We  now assume that this relation holds
at high redshift as well.

We parameterize  the mean LV relation as:  $M_B= a(z)  - b(\log 2\vm -
2.5) +  5\log  h$, where   only the   zero-point may   evolve in  time
according to the parametric functional form
\begin{equation}
a(z)=a_0 -2.5 \log F(z), \quad F(z) = (1+z^\alpha)^{\beta/\alpha} .
\end{equation}
We choose $a_0 = -18.71$  and $b=6.76$, corresponding to the empirical
local  LV relation used  in G99.  The  parameters $\alpha$ and $\beta$
allow a  wide range of shapes, where  $\alpha$ governs the behavior at
$z<1$,   and  $\beta$ the  asymptotic  shape at  larger redshifts. The
indications for   only little evolution   at $z\lsim  1$   guide us to
consider mostly large positive values for $\alpha$, while the value of
$\beta$ is unconstrained, and even its sign is unknown.

In Figure 2 we consider five different representative cases for the LV
zero-point evolution $a(z)$,  and show the  predicted evolution of the
number density of  galaxies with $M_B \leq  -20.2$, relative to  their
number density at $z=0$.   The case $(\alpha,\beta)=(0,0)$ is the case
of no evolution in the LV relation.  The case $(2,-1)$ is close to the
evolution predicted for the \lcdm\ model if the mass-to-light ratio is
constant in time.  In this model, because  halos  of fixed $\vm$  have
smaller masses at high redshift,  they  are fainter and therefore  the
predicted number of bright galaxies decreases.

The  case $(1,1)$ is close to  the prediction of a semi-analytic model
(the ``accelerated quiescent'' model  of Somerville, Primack, \& Faber
2000) in  which the efficiency   of star  formation scales  with   the
inverse dynamical  time of the halo. This  scaling is motivated by one
of the empirical   recipes    suggested Kennicutt (1998),    based  on
observations of nearby  galaxies.  The dynamical time  for a halo of a
fixed  velocity  is  smaller at  high  redshift   (because  of the  higher
density), and so galaxies of a fixed velocity are
\emph{brighter}.
This model predicts an
\emph{increase} in the number density of bright galaxies out to
$z\sim5$. Although this star formation recipe is  a standard choice in
semi-analytic models  (e.g.,  Kauffmann, White, \&  Guiderdoni  1993),
note that it predicts fairly strong  evolution in the LV relation even
at $z\sim1$. The other two  cases, $(5,1)$ and $(4,-4)$, are arbitrary
choices that illustrate a  wide  range of possibilities while  keeping
the LV relation almost unchanged out to $z\sim 1$.

{\pspicture(0,0)(11,11.)
\rput[tl]{0}(-.1,11.){\epsfxsize=8.9cm
\epsffile{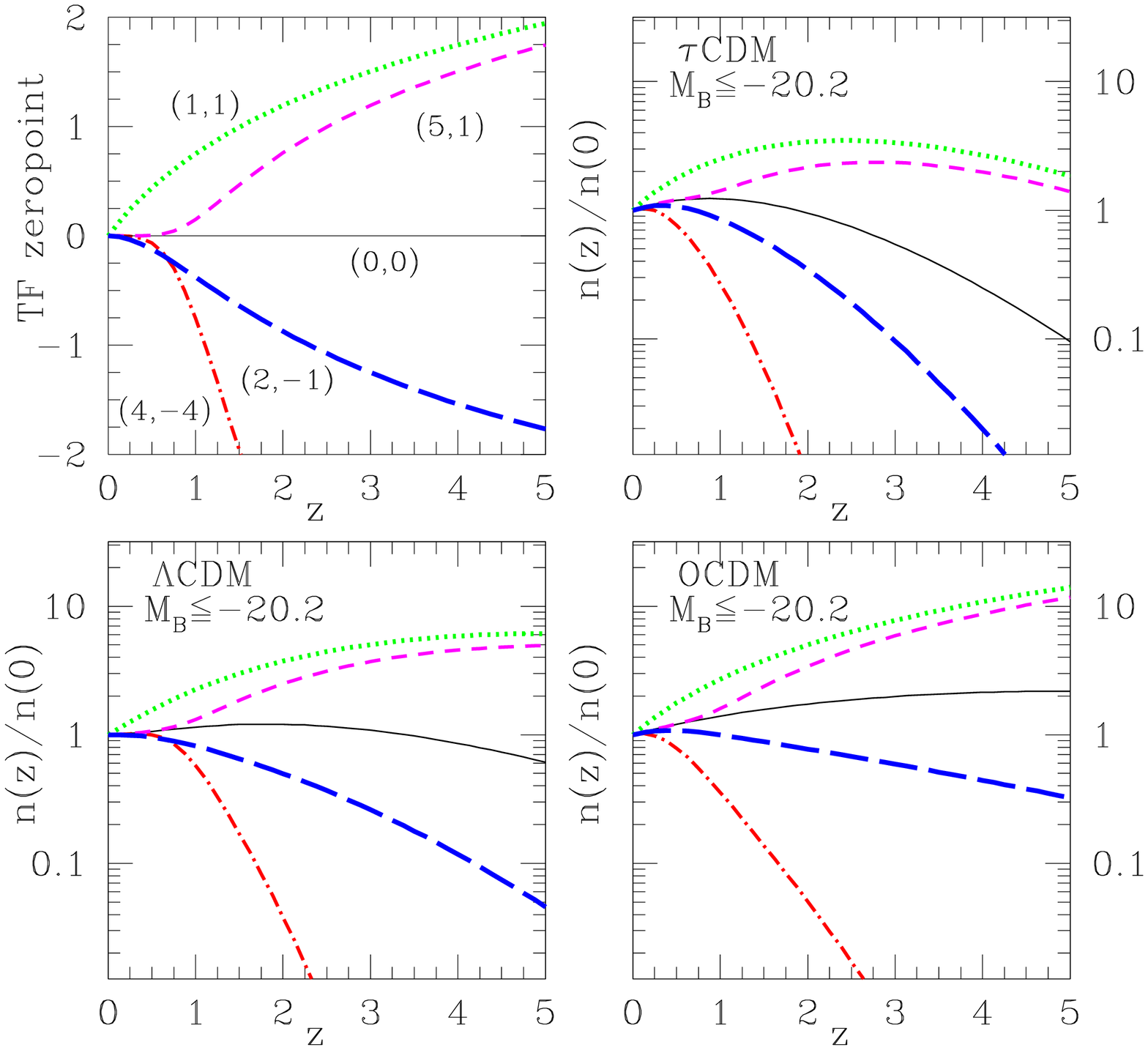}}
\rput[tl]{0}(0,2.){
\begin{minipage}{8.5cm}
  \small\parindent=4.5mm   {\sc Fig.}~2.---  Five parameterized models
  marked by  $(\alpha,\beta)$ for the evolution  of the LV zero-point,
  $a_0-a(z)$ (Eq.~1; larger  values   are brighter), along  with   the
  implied relative evolution    of the comoving  density of   galaxies
  brighter  than  $M_B =  -20.2$  for  the three cosmological  models.
  {\vskip 0.5 cm}

\end{minipage}
}
\endpspicture}

\section{Discussion}
\label{sec:conc}

We have  argued  that observed  estimates  of  the  number density  of
galaxies above a  given luminosity as  a  function of redshift can  be
used to constrain the redshift evolution of the  LV relation. Lilly et
al. (1995) found only little evolution in the number density of
\emph{bright} galaxies in the rest-frame B-band out to $z\sim1$,
consistent with  the direct constraints  on  LV evolution out to  this
redshift (Vogt et al. 1996, 1997,  2000). Redshift surveys of the past
decade were  unable to reach  the higher  redshift range  $1.0 \lsim z
\lsim  2.0$ associated with the spectroscopic  'desert' or `high place
of sacrifice'.   The results presented  by  Dickinson (2000), based on
NICMOS (near-IR) imaging  of the Hubble Deep  Field (HDF) suggest that
the assembly  of  present-day luminous galaxies  may   be taking place
precisely in this   redshift interval.   Using photometric  redshifts,
Dickinson  finds that there are  only  one-third as many galaxies with
rest-frame $M_V < -19$ in the redshift interval  $1.4 \lsim z \lsim 2$
as  in the interval  $0  \lsim z  \lsim  1.4$. In our  fiducial \lcdm\
cosmology, the comoving volumes of the HDF for these two intervals are
equal, implying an  overall drop in the comoving  number density at $z
\ga 1.4$ of about a factor of three.   Taken at face value, this would
correspond to  dramatic evolution  in the  LV   relation at  $z\ga 1$,
rather similar to the most extreme `dimming' model shown in Figure~2.

Probably the weakest  link in our  calculation is our assumption  that
the halo  maximum velocity   is the same    as the  measured  internal
velocity for a galaxy. The actual measured velocity will depend on the
baryon fraction   in  the galaxy,   the   spatial distribution of  the
baryons, and the degree to which angular  momentum is preserved in the
collapse. However, note that because we expressed the space density in
Figure~2 \emph{relative}  to that at  $z=0$, our plotted  results will
not change by much as  long as the ratio of  observed velocity to  the
halo maximum velocity,  $C \equiv v_{\rm  obs}/\vm$, is of order unity
and  roughly constant with redshift.   We have calculated the value of
$C(z)$ using a simple analytic model  of disk formation, which assumes
that a constant fraction  of the halo mass condenses  into a disk, and
angular momentum  is conserved (Mo, Mao,\&  White 1998).  We find that
$C(z) \simeq 1.3$ for  the  redshift range $0   \lsim z \lsim 2$,  and
falls  to $C(z)  \simeq 1.0$  by  $z\sim5$.   This suggests  that  our
predictions should be reasonably robust at least out to $z\sim2$.

How  significant is the  effect seen in   the HDF?  One possibility is
that galaxies are preferentially  missed at $z\ga  1.4$ because of the
$(1+z)^4$   cosmological  surface-brightness dimming. Dickinson (2000)
claims that  this is unlikely to be  a  large effect; when  bright HDF
galaxies from $z\lsim1$ are artificially redshifted to $z\lsim 2$ with
no  intrinsic  luminosity evolution,  most of   them  are still easily
detectable. Another possibility  is that the photometric redshifts are
inaccurate; Dickinson points out that there are
\emph{no} spectroscopic redshifts in the range $1.4\lsim z\lsim 2$ where
the  deficit  is seen. However, this   would require quite  a dramatic
failure where \emph{all}  of the galaxies (with photometric redshifts)
assigned    to  $0.5\lsim z_{\rm phot}\lsim  1.4$    would have  to be
reassigned   to the higher  redshift  range.  Most importantly, recall
that the  HDF covers a relatively  small volume,  and therefore random
fluctuations  along the  line of  sight are likely  to  be large.  The
observed deficit   of   galaxies   might simply   reflect  large-scale
``spikes'' in the redshift distribution.
Although there  is  some evidence for   a  similar deficit  of  bright
galaxies at $z  \gsim 1$ in other deep  optical/IR surveys (Fontana et
al. 1999),  secure confirmation  of these  results  will require  deep
wide-field surveys with multi-band optical-IR photometry. Several such
surveys are now in  progress (e.g. the Cambridge-Carnegie Las Campanas
IR Survey).  Moreover,  the DEEP survey  (Davis  \&  Faber 1998)  will
obtain measured  line-widths for a  large number of  galaxies at $z\ga
0.7$, which will provide direct constraints on the LV relation at high
redshift.

Finally, our results may be relevant to a recent proposal by Newman \&
Davis (2000) to use  velocity-selected galaxies as a tracer population
in a classical   $dN/dz$  measure of  cosmological parameters.    This
analysis  would be subject to   similar uncertainties in relating  the
halo velocity to  observable galaxy velocity,  e.g., the dependence on
whether the halos are selected by $\vm$ or $\vvir$.  Moreover, because
the  proposed    spectroscopic surveys  are   magnitude  limited,  the
indicated evolution   of   the  LV  relation could    introduce severe
selection biases into the analysis.

\section*{Acknowledgments}
\begin{small}
This work has  been performed at the ITP,  Santa Barbara.  It has been
supported by  grants from the US-Israel BSF,  the  Israel SF, NASA and
NSF at UCSC and NMSU.  JSB was supported  by NASA LTSA grant NAG5-3525
and NSF grant  AST-9802568.  JRP acknowledges  a Humboldt Award at the
Max Planck  Institute for Physics  in Munich.  RSS  is supported  by a
rolling grant from PPARC.  We thank Marc  Davis and Mark Dickinson for
helpful correspondence.
\end{small}

\def\re{\reference}

\end{document}